\documentclass[byrevtex,superscriptaddress,onecolumn,preprint]{revtex4}%
\usepackage{amsfonts}
\usepackage{amsmath}
\usepackage{amssymb}
\usepackage{graphicx}%
\providecommand{\U}[1]{\protect\rule{.1in}{.1in}}
\begin{document}
\preprint{ }
\title{Self-Similar Modes of Coherent Diffusion}
\author{O. Firstenberg}
\affiliation{Department of Physics, Technion-Israel Institute of Technology, Haifa 32000, Israel}
\author{P. London}
\affiliation{Department of Physics, Technion-Israel Institute of Technology, Haifa 32000, Israel}
\author{D. Yankelev}
\affiliation{Department of Physics, Technion-Israel Institute of Technology, Haifa 32000, Israel}
\author{R. Pugatch}
\affiliation{Department of Physics of Complex Systems, Weizmann Institute of Science,
Rehovot 76100, Israel}
\author{M. Shuker}
\affiliation{Department of Physics, Technion-Israel Institute of Technology, Haifa 32000, Israel}
\author{N. Davidson}
\affiliation{Department of Physics of Complex Systems, Weizmann Institute of Science,
Rehovot 76100, Israel}

\pacs{42.50.Gy, 42.25.Fx}

\begin{abstract}
Self-similar solutions of the coherent diffusion equation are derived and
measured. The set of real similarity solutions is generalized by the
introduction of a nonuniform phase surface, based on the elegant Gaussian
modes of optical diffraction. In an experiment of light storage in a gas of
diffusing atoms, a complex initial condition is imprinted, and its diffusion
dynamics is monitored. The self-similarity of both the amplitude and the phase
pattern is demonstrated, and an algebraic decay associated with the mode order
is measured. Notably, as opposed to a regular diffusion spreading, a
self-similar contraction of a special subset of the solutions is predicted and observed.

\end{abstract}
\maketitle

Self-similar solutions are generally associated with the long-time behavior of
dynamic processes \cite{Barenblatt1996} and found in nearly all disciplines,
from astrophysics and fluid dynamics to condensed matter and optics
\cite{GoldmanRMP1984,ColemanPRL1994,WadatiJPSJ1998,HarveyPRL2003,vanHeijstPOF2004,WisePRL2004}%
. In most dissipative systems, similarity solutions decay with a
characteristic rate, indicating the asymptotic evolution of a given initial
condition \cite{KloosterzielJEM1990,vanHeijstPOF2004}. Similarity solutions
emerge also in non-dissipative systems and often prevail, with a familiar
example being the family of Gaussian beams in free-space paraxial optics
\cite{SiegmanBook}, when the propagation distance is given the role of time.
Gaussian beams are broadly referred to as \emph{modes }of diffraction, even
though they are not nondiffracting as Bessel beams \cite{DurninPRL1987} and
not eigenmodes of an underlying Hamiltonian. Moreover, they are only partially
self-similar throughout the propagation -- their shape is preserved up to
scaling and normalization, while their phase pattern curves or flattens.

The imaginary-time counterpart of paraxial diffraction is the coherent
diffusion of a complex-valued field,%
\begin{equation}
\frac{\partial\psi}{\partial t}=D\nabla_{\perp}^{2}\psi-\gamma\psi,
\label{Eq_diffusion}%
\end{equation}
with $\psi=\psi(x,y;t)$ being a two-dimensional coherence field, $D$ a real
coefficient, and $\gamma$ a linear decay rate. Coherent diffusion in the form
of Eq. (\ref{Eq_diffusion}) arise, for example, in the thermal motion of hot
coherent atoms, where the field of internal-state quantum coherence ($\psi$)
is subjected to atomic diffusion ($D$) and decoherence ($\gamma$)
\cite{FirstenbergPRA2008}. Using the technique of light storage and retrieval
\cite{PhillipsPRL2001}, any arbitrary initial condition can be imprinted on
the diffusing atoms, and the subsequent dynamics can be observed
\cite{ShukerImaging2007}. This system attracted considerable recent study,
exhibiting spectral fringes, narrowing, and coherent recurrence
\cite{ZibrovPRA2001,WalsworthPRL2006,PugatchPRL2009}, magnetization diffusion
\cite{WeisOL2000}, diffusion of vortices \cite{PugatchPRL2007}, transverse
momentum diffusion \cite{YelinPRA2008,HowellPRL2008}, and slow-light spreading
\cite{FirstenbergNatPhys2009}.

In this letter, we present and explore the exact self-similar complex
solutions of Eq. (\ref{Eq_diffusion}). We experimentally create and follow the
dynamics of several self-similar modes and demonstrate the preservation of
both their shape and phase pattern, as well as their characteristic decay. A
self-similar contraction, resembling focus or collapse \cite{GoldmanRMP1984},
is also demonstrated.%
\begin{figure}
[ptb]
\begin{center}
\includegraphics[
height=4.815cm,
width=7.183cm
]%
{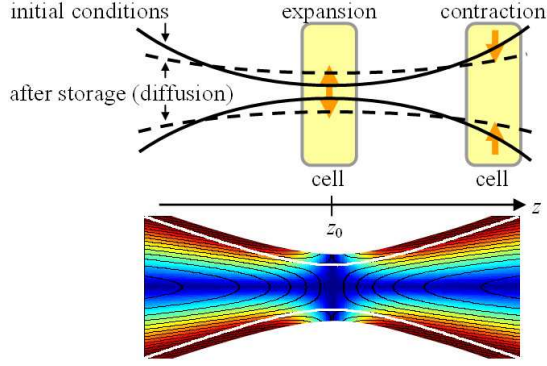}%
\caption{Top: the effect of diffusion on a Gaussian beam is an effective
stretching of the waist radius, even when the diffusion does not occur at the
waist plane. Therefore, far enough from the waist, the diffusion results in
the contraction of the transverse shape. Bottom: colormap of the phase
gradients of the field envelope (black lines are equal phase contours; white
lines are the beam outline). From the viewpoint of the microscopic atomic
motion, the contraction far from the waist occurs due to destructive
interference of atoms diffusing through the rapidly oscillating phase pattern
(red colored).}%
\label{fig_out_of_waist_sketch}%
\end{center}
\end{figure}

Consider first the time-independent paraxial diffraction, $\partial E/\partial
z=-i\nabla_{\perp}^{2}E/(2k)$, for the slowly-varying envelope, $E$, of a
light field with a wave number $k.$ Two different sets of polynomial-Gaussian
solutions are known for this equation, namely the standard and the\ elegant
beams \cite{SiegmanBook}. The more familiar 'standard' modes, \textit{e.g.}
the Hermite-Gaussian (HG) or Laguerre-Gaussian (LG), form complete sets of
modes that are self-similar under diffraction. Their transverse intensity
distribution, $I\left(  x,y;z\right)  $, is maintained along the propagation
direction $z$, normalized, and scaled by the beam radius $w(z)$. In contrast,
the transverse shape of the 'elegant' solutions is generally not maintained,
and originally they were investigated due to their elegant mathematical form
\cite{SiegmanJOSA1973,WunscheJOSA1989}.

The elegant HG solution, $E_{n_{1},n_{2}}^{\text{HG}}(\mathbf{r};w_{0}),$ with
$w_{0}$ being the radius at the waist plane $z=0$, is written in terms of the
Hermite polynomials of orders $n_{1}$ and $n_{2}$ as,%
\begin{equation}
E_{n_{1},n_{2}}^{\text{HG}}=\frac{E_{0}}{kw_{0}}\left[  \frac{kw_{0}^{2}%
}{2q\left(  z;w_{0}\right)  }\right]  ^{\frac{N+2}{2}}H_{n_{1}}\left(
\tilde{x}\right)  H_{n_{2}}\left(  \tilde{y}\right)  e^{-\tilde{x}^{2}%
-\tilde{y}^{2}}. \label{eq_elegant_form}%
\end{equation}
Here, $q(z;w_{0})=iz_{R}+z$ is the complex radius, $z_{R}=kw_{0}^{2}/2$ is the
Rayleigh length, $N=n_{1}+n_{2}$ is the total mode order, and $E_{0}$ is a
normalization constant. The transverse scaling, appearing in both the
polynomial and the Gaussian terms, depends on $z$ and $w_{0}$, with $\tilde
{x}=x[ik/2/q(z;w_{0})]^{1/2}$ and $\tilde{y}=y[ik/2/q(z;w_{0})]^{1/2}$. The
beam radius, $w(z),$ and the radius of curvature of the phase fronts, $R(z),$
are obtained from $q(z;w_{0})^{-1}=R(z)^{-1}-(i2/k)w(z)^{-2}.$ For the
corresponding standard mode, the complex arguments of the polynomial
($\tilde{x}$, $\tilde{y}$) are replaced by real arguments [$\sqrt{2}x/w(z)$,
$\sqrt{2}y/w(z)$]. Thus elegant modes with a homogeneous polynomial are also
standard, and we denote them as 'common' modes.

Now suppose that a two-dimensional diffusion takes place at a certain $(x,y)$
plane, where $z$ is held constant. Eq. (\ref{eq_elegant_form}), with $\partial
E/\partial z=-i\nabla_{\perp}^{2}E/(2k)$ and $(\partial/\partial w_{0}%
)_{z}=ikw_{0}(\partial/\partial q)_{w_{0}}+(\partial/\partial w_{0})_{q},$
then gives
\begin{equation}
\nabla_{\perp}^{2}E_{n_{1},n_{2}}^{\text{HG}}=\frac{2}{w_{0}}\left(
\frac{\partial E_{n_{1},n_{2}}^{\text{HG}}}{\partial w_{0}}\right)  _{z}%
-\frac{2}{w_{0}^{2}}\left(  N+1\right)  E_{n_{1},n_{2}}^{\text{HG}}.
\label{eq_laplacian}%
\end{equation}
Under diffusion, the first term in the right-hand side of Eq.
(\ref{eq_laplacian}) accounts for a change in the waist radius, and the second
term for an homogenous decay of the field. Eq. (\ref{Eq_diffusion}) is
therefore solved by%
\begin{equation}
\psi_{n_{1},n_{2}}^{\text{HG},z}\left(  x,y,t\right)  =e^{-\gamma t}s\left(
t\right)  ^{-\left(  N+1\right)  }E_{n_{1},n_{2}}^{\text{HG}}\left(
\mathbf{r};w_{0}s\left(  t\right)  \right)  , \label{eq_psi_nm}%
\end{equation}
where the diffusion coefficient enters only through the waist stretching
factor, $s\left(  t\right)  =(1+4Dt/w_{0}^{2})^{1/2}$. Thus the spatial
consequence of diffusion is always an effective stretching of the beam radius
\emph{at the waist plane}, even when the diffusion occurs far from the waist
($z\neq0$), as illustrated in Fig. \ref{fig_out_of_waist_sketch} (top). Note
that if diffusion is addressed as an imaginary-time evolution of diffraction,
it is readily seen from the definition of the complex radius $q(z;w_{0})$ that
exchanging the real evolution in $z$ for an imaginary one corresponds to a
real increase in the waist radius. The total power, $P(t)=\int dxdy|\psi
_{n_{1},n_{2}}^{\text{HG},z}|^{2},$ which is independent of $z,$ is not
preserved under diffusion even when $\gamma=0,$ due to an algebraic decay term
$s\left(  t\right)  ^{-(N+1)}$. This occurs even for the lowest Gaussian mode
($N=0$) because it is the field, rather than the intensity, that is diffusing.
Higher-order modes ($N>0$) decay faster due to the diffusion of the
non-homogenous phase pattern, which contains larger gradients for higher $N$.
A similar procedure can be carried out for an elegant LG solution,
$E_{p,m}^{\text{LG}}$, of radial order $p$ and orbital order $m$, yielding Eq.
(\ref{eq_psi_nm}) for the diffusing field $\psi_{p,m}^{\text{LG},z}(x,y,t),$
with $N=p+m$.%

\begin{figure}
[ptb]
\begin{center}
\includegraphics[
height=12.6855cm,
width=8.1012cm
]%
{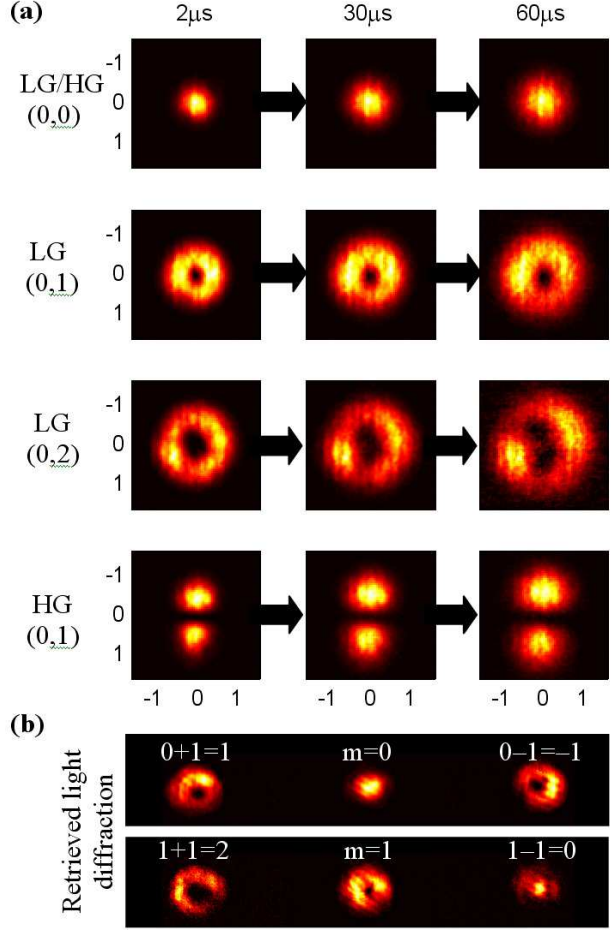}%
\caption[Diffusion of the complex field of atomic coherence, demonstrating the
self-similarity property.]{(a) Diffusion of the complex field of atomic
coherence in a storage-of-light experiment, demonstrating the self-similarity
property of (top to bottom): the basic Gaussian mode, the Laguerre-Gaussian
modes with radial order $p=0$ and orbital orders $m=1,2,$ and the
Hermite-Gaussian mode with cartesian orders $(0,1).$ All images are
$1.6\times1.6$ mm. (b) Images after storage, diffracted by a binary grating
mask with a fork dislocation, for confirming the conservation of phase. The
retrieved Gaussian mode (top) yields the two $m=\pm1$ vortex modes, while the
retrieved vortex $m=+1$ (bottom) produces an $m=0$ and an $m=+2$ modes.}%
\label{fig_pictures}%
\end{center}
\end{figure}

At the waist, all arguments in Eq. (\ref{eq_elegant_form}) are real, and the
HG solutions $\psi_{n_{1},n_{2}}^{\text{HG},z=0}$ are identified with the
expanding similarity solutions of regular diffusion \cite{KloosterzielJEM1990}%
, occurring, \textit{e.g.}, for the vorticity field of a viscous fluid
\cite{KloosterzielJFM1991,vanHeijstPOF2004}. The real solutions are
alternatively derived from a given self-similar solution with a single real
scaling ($\psi(\mathbf{r},t)=$ $h(t)f[\mathbf{r}/w(t)]$), by taking any of its
spatial derivatives ($\partial^{n}\psi/\partial x^{n}=h(t)f^{(n)}%
[\mathbf{r}/w(t)]/w(t)^{n}$). Indeed, the derivatives of the lowest-order
Gaussian beam constitute the elegant modes, with the derivative order
corresponding to the total mode order $N,$ and with the possible
generalization for unified Hermite-Laguerre-Gaussian modes
\cite{GuoOL2008,ZaudererJOSA1986}. Consequently, $\psi_{n_{1},n_{2}%
}^{\text{HG},z=0}$ or $\psi_{p,m}^{\text{LG},z=0},$ and all linear
combinations of them with the same total order, are self-similar, sharing the
same stretching and the same algebraic decay. Alternatively, any complex
'image' in the $(x,y)$ plane can be expanded in terms of $\psi_{n_{1},n_{2}%
}^{\text{HG},z=0}$ or $\psi_{p,m}^{\text{LG},z=0}$ using their biorthogonal
pairs \cite{SiegmanJOSA1997}, its diffusion can be described in terms of the
modes dynamics, and asymptotically the lowest-order solution prevails
\footnote{Note that \emph{integrals} of the fundamental Gaussian are also
self-similar and may be employed in the expansion of solutions with infinite
energy.}.

To validate the above predictions, we perform experiments with thermal alkali
atoms confined in a vapor cell with a buffer gas. A resonant laser beam with
the desired optical mode is sent into the cell, and its complex field envelope
is mapped onto the atomic coherence field utilizing a storage-of-light
technique, by shutting down an auxiliary control beam
\cite{PhillipsPRL2001,ShukerImaging2007}. The coherence field, $\psi$, is
allowed to evolve for a controlled duration $\tau,$ in which the alkali
diffusion through the buffer gas takes place, as well as a homogenous
decoherence (\emph{e.g.} due to spin-exchange relaxation \cite{ShukerPRA2008}%
). The coherence is then converted back to light, which is imaged onto a
camera. The experimental set-up and procedure is similar to that described in
\cite{PugatchPRL2007}, where the topological stability of the vortex in a
stored $E_{0,1}^{\text{LG}}$ mode was attested.

\begin{figure}
[ptb]
\begin{center}
\includegraphics[
height=8.044cm,
width=8.211cm
]%
{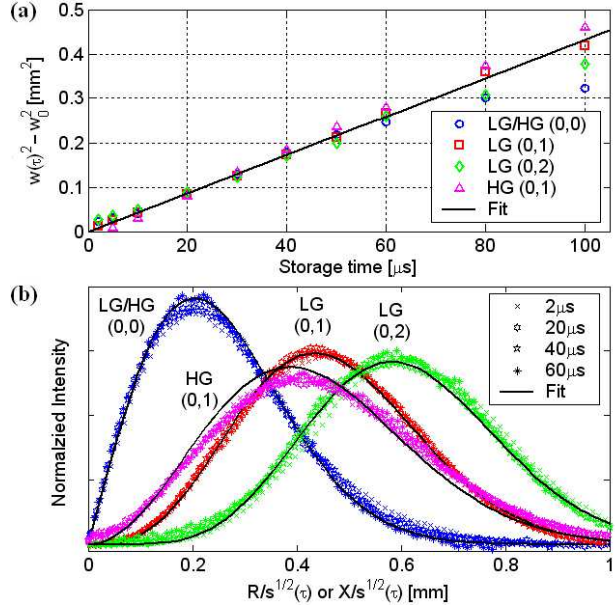}%
\caption{(a) Linear growth of $w^{2}$ with respect to the time duration of
diffusion. The line is $w^{2}-w_{0}^{2}=4D\tau$, with $D=10.8$ cm$^{2}/$s.
Both $D$ and $w_{0}$ were fit parameters ($w_{0}=0.4\sim0.55$ mm, varying for
the different modes). (b) Self-similarity: cross-sections (cartesian or
radially-weighted) at different times are congruent when plotted versus the
scaled coordinate [$\tilde{y}(\tau)=s(\tau)^{-1/2}y$ for the cartesian HG
mode, $\tilde{r}(\tau)=s(\tau)^{-1/2}r$ for the cylindrical LG modes]. The
solid lines are the analytic forms of the elegant HG and LG modes.}%
\label{fig_waists}%
\end{center}
\end{figure}

\begin{figure}
[ptb]
\begin{center}
\includegraphics[
height=2.4647in,
width=3.1877in
]%
{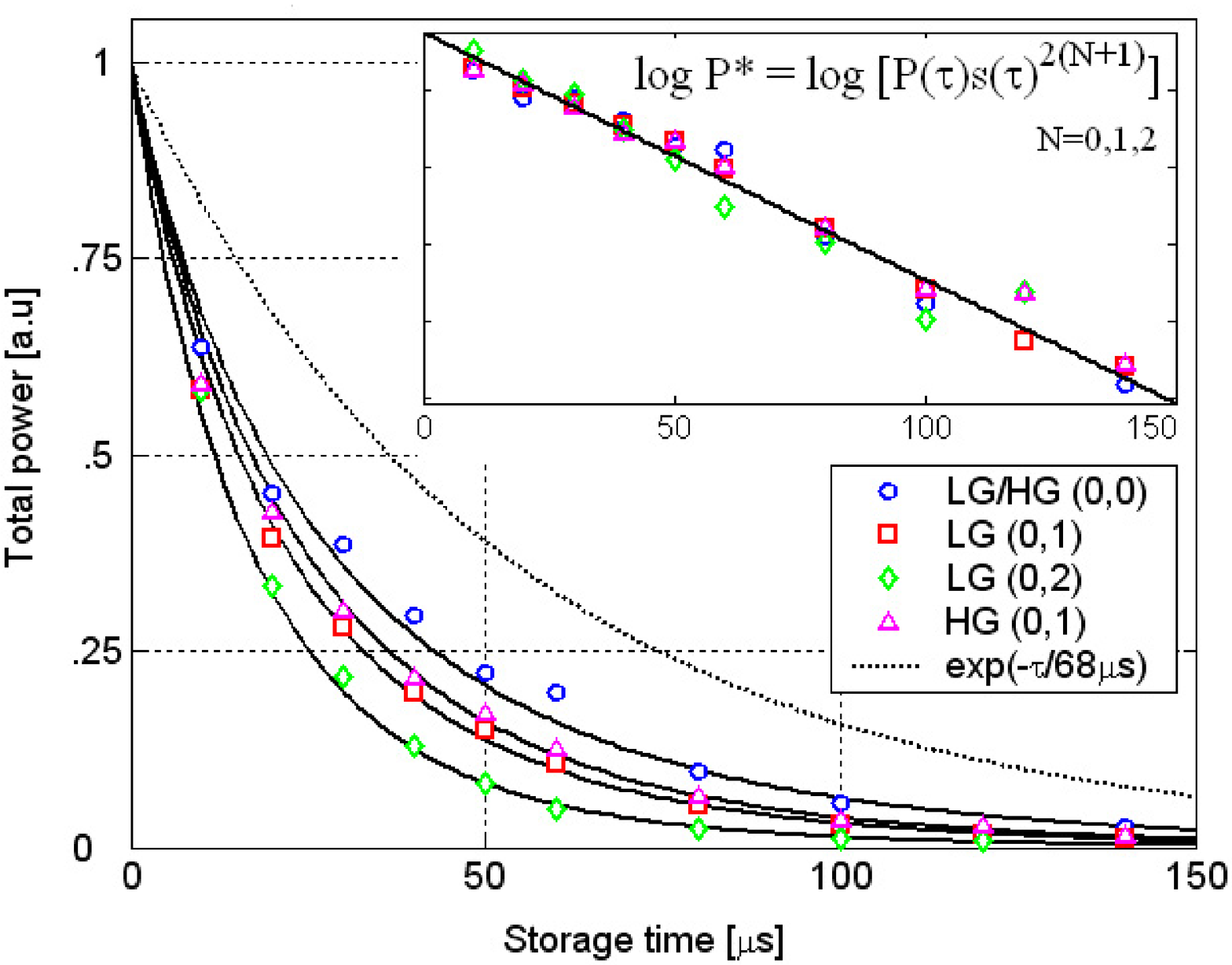}%
\caption[Decay of the total power, obtained from the direct integration over
the retrieved images.]{Decay of the total power in the retrieved images
$P(\tau)=\int dxdy|\psi(\tau)|^{2}$. In the inset, $P^{\ast}$ compensates for
the algebraic decay [$s(\tau)$ known from Fig. \ref{fig_waists}] and collapses
onto\ a single straight line in the semilog scale, yielding the homogenous
decay rate $2\gamma=(68$ $\mu$s$)^{-1}.$ In the main graph, the dashed line is
$e^{-2\gamma\tau}$ and the solid lines are $e^{-2\gamma\tau}s(\tau
)^{-2(N+1)},$ demonstrating the faster decay of the higher-order modes. The
difference between the LG$_{0,1}$ and the HG$_{0,1}$, both having $N=1$, is
due to slightly different initial waist radii ($w_{0}$).}%
\label{fig_powers}%
\end{center}
\end{figure}

The experiment was carried out with the fundamental Gaussian mode
$E_{0,0}^{\text{(HG/LG)}}$, the LG modes $E_{0,1}^{\text{LG}}$ and
$E_{0,2}^{\text{LG}},$ and the HG mode $E_{0,1}^{\text{HG}}.$ Fig.
\ref{fig_pictures}(a) presents the retrieved images, proportional to
$|\psi(x,y,\tau)|^{2},$ for a storage performed with the cell located at the
beam waist ($z=0$), for durations of $\tau=$ $2,~30$, and $60$ $\mu$s.
Evidently, all the modes expand but maintain their shape through the diffusion
process. As a complementary test, we have also passed the retrieved beams
through a binary grating mask with a fork dislocation, which adds a phase
function $m\phi$ in its $m$-th diffraction order \cite{SoskinOptComm1993}.
After the mask, as shown in Fig. \ref{fig_pictures}(b), the retrieved vortex
mode $E_{0,1}^{\text{LG}}$ ($m=1$) produces a Gaussian ($m=0$) and a
higher-order vortex ($m=2$) in the $-1$ and $+1$ diffraction orders,
confirming the maintenance of the phase pattern. Fig. \ref{fig_waists}(a)
presents the increase in the waist-radii squared versus the storage duration,
showing the same linear increase for all curves, $w(\tau)^{2}-w_{0}^{2}%
=w_{0}^{2}[s\left(  \tau\right)  ^{2}-1]=4D\tau$. The cross-sections shown in
Fig. \ref{fig_waists}(b), scaled according to $s\left(  \tau\right)  $ and
normalized, clearly demonstrate the self-similarity. The algebraic decay of
the diffusing modes, $s\left(  t\right)  ^{-(N+1)},$ is measured by
integrating over the intensity of the retrieved images (Fig. \ref{fig_powers}%
). All modes exhibit a significant algebraic decay on top of the homogenous
decay, with the higher-order modes decaying faster, showing an excellent
quantitative agreement with the predictions of Eq.(\ref{eq_psi_nm}).%

We now discuss the diffusion of the elegant modes at a given plane outside the
waist plane, $z\neq0$. For Gaussian modes, an expansion of the waist radius
results in an increase of the beam's transverse size for $|z|<z_{R}$ and in a
decrease for $|z|>z_{R}$ (Fig. \ref{fig_out_of_waist_sketch}, top). It
follows, perhaps counterintuitively, that the initial effect of diffusion
occurring at $|z|>z_{R}$ is a contraction. Locally, it is the consequence of a
destructive interference in regions of the beam where the phase pattern
rapidly oscillates (Fig. \ref{fig_out_of_waist_sketch}, bottom). The
contraction versus storage time, $C(\tau)=w(\tau)/w(0),$ is given by
$C(\tau)^{2}=[s(t)^{4}+\rho^{2}]/[s(t)^{2}(1+\rho^{2})],$ where $\rho=z/z_{R}$
specifies the initial distance from the waist. As the waist radius increases
during the diffusion, $z_{R}(\tau)$ increases and eventually crosses the
observed plane (which $z$ coordinate is constant). At this time, the maximal
contraction $C_{\min}^{2}=2\rho/(1+\rho^{2})$ is obtained, and thenceforth the
beam expands indefinitely.%

\begin{figure}
[ptb]
\begin{center}
\includegraphics[
height=5.4366cm,
width=7.9474cm
]%
{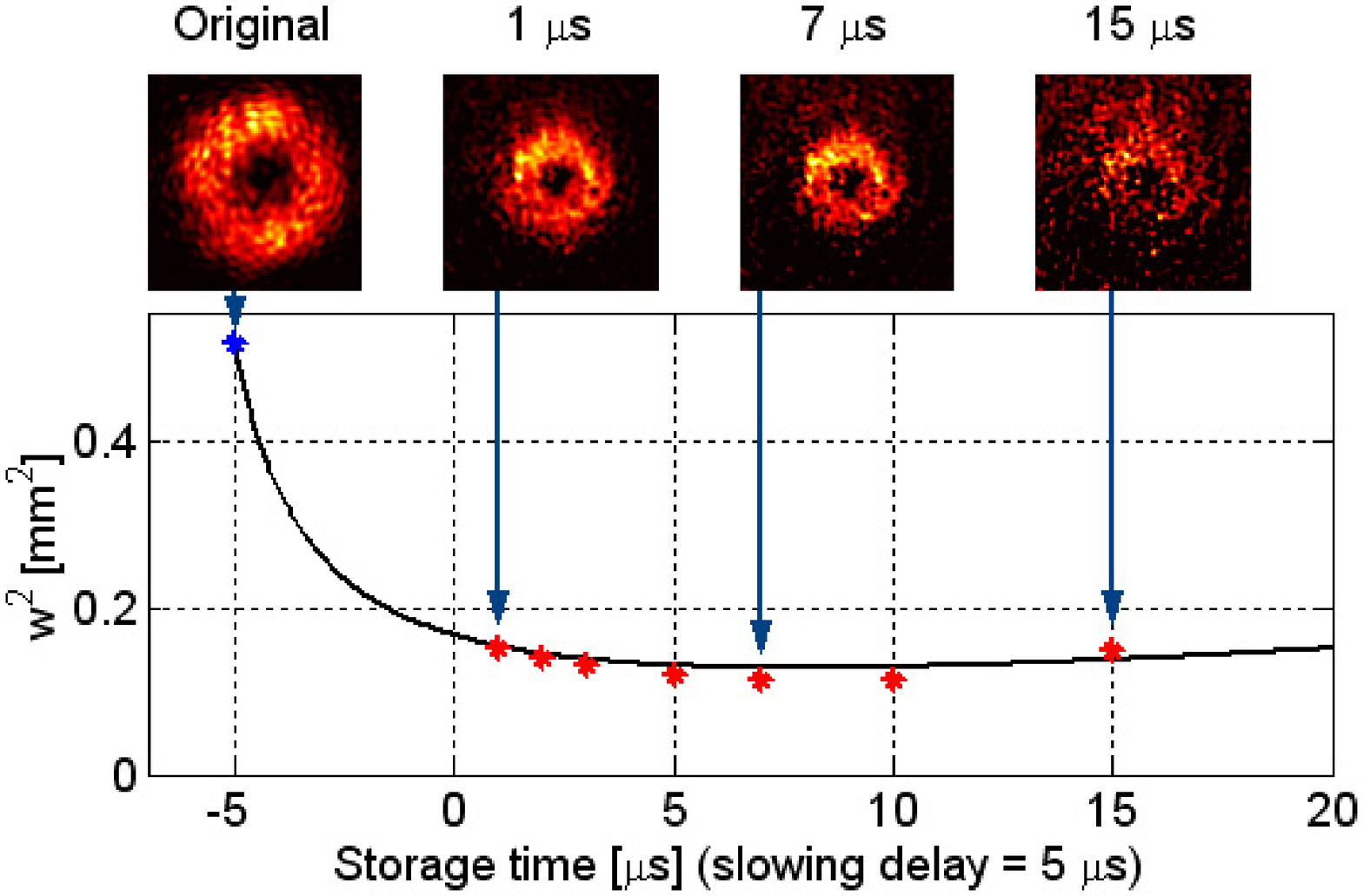}%
\caption{The self-similar contraction and subsequent expansion of the
LG$_{0,1}$ mode upon diffusion. Contrary to the former experiments, here we
refer to the original beam as the initial condition (left, taken
off-resonantly), since much contraction occurs already during the slow-light
propagation in cell. The original mode has a radius of $700$ $\mu$m and a
phase curvature of ($250$ mm)$^{-1}$. The minimal radius, of $340$ $\mu$m, is
obtained after slowing for $5$ $\mu$s plus storage for $7$ $\mu$s. The black
line is calculated from $C(\tau)^{2}$ with no fit parameters. In these
conditions, the power decays substantially faster than in the expansion
experiments, and substantial noise is already apparent after $15$ $\mu$s of
storage.}%
\label{fig_shrink}%
\end{center}
\end{figure}

Elegant modes, as opposed to the standard modes, are generally not
self-similar under \emph{diffraction}, and the shape of the beam depends on
$z/z_{R}$. Hence, even when $z$ is held constant, the increase of $z_{R}$
during \emph{diffusion} changes the transverse shape and breaks its
self-similarity. However, the aforementioned 'common' modes, which are
simultaneously elegant and standard, are self-similar under diffraction and
thus also self-similar under diffusion even for $z\neq0$. The HG modes of
orders $0$ and $1$ \cite{SiegmanBook} and all LG modes with $p=0$ (the vortex
modes) are such common modes. Far from the waist, at $z>z_{R}$, these modes
contract self-similarly. Fig. \ref{fig_shrink} presents the experimental
result for a diffusing $E_{0,1}^{\text{LG}}$ mode focused at a distance
$8z_{R}$ before the cell ($\rho=8$), yielding $C_{\min}^{2}\approx1/4.$

Finally, we point out an intriguing instability phenomenon noticeable in Fig.
\ref{fig_pictures}(a) for the $E_{0,2}^{\text{LG}}$ mode. During diffusion,
the $m=2$ vortex breaks-down into two vortices, probably of $m=1$. A decay of
high-order vortices into lower-order ones has been seen also in optics,
quantum fluids, and Bose-Einstein condensates. Here, several candidate
mechanisms may be responsible for the imperfection
\cite{KloosterzielJFM1991,SoskinOptComm1993}, which evidently conserves the
cross-section of the original vortex [Fig. \ref{fig_waists}(b)].

In conclusion, we have shown that when elegant Gaussian modes are put through
coherent diffusion, their waist radius effectively expands. The total power in
the field decays algebraically during the diffusion, even for the lowest-order
mode, due to field interference effects. The complete set of elegant modes is
self-similar at the waist, while far from the waist, self-similarity is found
only for the subset of common modes. For the latter, a self-similar
contraction may occur. The cross-sections at the waist of the standard and the
elegant Gaussian modes form a complete set of self-similar modes for
diffraction and diffusion, respectively. A similar set for the simultaneous
process of diffraction and diffusion, as occurring in the dynamics of slow
light \cite{FirstenbergNatPhys2009}, is yet to be explored.

We gratefully acknowledge discussions with A. Ron.

\bibliographystyle{apsrev}

\end{document}